# Charged Oxygen Vacancy Induced Ferroelectric Structure Transition in Hafnium Oxide

Ri He[1#], Hongyu Wu[1#], Shi Liu[2, 3, 4], Houfang Liu[5], X. Renshaw Wang[6, 7] and Zhicheng Zhong[1, 8*]

[1]Key Laboratory of Magnetic Materials Devices & Zhejiang Province Key Laboratory of Magnetic Materials and Application Technology, Ningbo Institute of Materials Technology and Engineering, Chinese Academy of Sciences, Ningbo 315201, China

[2]School of Science, Westlake University, Hangzhou, Zhejiang 310024, China

[3]Institute of Natural Sciences, Westlake Institute for Advanced Study, Hangzhou, Zhejiang 310024, China

[4]Key Laboratory for Quantum Materials of Zhejiang Province, Hangzhou Zhejiang 310024, China

[5]Institute of Microelectronics and Beijing National Research Center for Information Science and Technology (BNRist), Tsinghua University, Beijing 100084 China

[6]Division of Physics and Applied Physics, School of Physical and Mathematical Sciences, Nanyang Technological University, 21 Nanyang Link, 637371 Singapore

[7]School of Electrical and Electronic Engineering, Nanyang Technological University, 50 Nanyang Ave, 639798 Singapore

[8]China Center of Materials Science and Optoelectronics Engineering, University of Chinese Academy of Sciences, Beijing 100049, China

**Abstract**

The discovery of ferroelectric $HfO_2$ in thin films and more recently in bulk is an important breakthrough because of its silicon-compatibility and unexpectedly persistent polarization at low dimensions, but the origin of its ferroelectricity is still under debate. The stabilization of the metastable polar orthorhombic phase was often considered as the cumulative result of various extrinsic factors such as stress, grain boundary, and oxygen vacancies as well as phase transition kinetics during the annealing process. We propose a novel mechanism to stabilize the polar orthorhombic phase over the nonpolar monoclinic phase that is the bulk ground state. Our first-principles calculations demonstrate that the doubly positively charged oxygen vacancy, an overlooked defect but commonly presented in binary oxides, is critical for the stabilization of ferroelectric phase. The charge state of oxygen vacancy serves as a new degree of freedom to control the thermodynamic stability of competing phases of wide-band-gap oxides.

---

[#] These authors contribute equally to this work
[*] zhong@nimte.ac.cn

The discovery of unusual ferroelectricity in fluorite-structural hafnium oxide ($HfO_2$) has attracted considerable attention because of its silicon-compatibility and unexpectedly robust ferroelectricity at low dimensions compared to conventional perovskite-based ferroelectrics [1-7]. But there is still an important issue unresolved regarding the origin of ferroelectricity, that is, the stability puzzle: the polar orthorhombic (PO) $Pca2_1$ phase is ferroelectric but metastable[8-10], while the nonpolar monoclinic (M) phase is the ground state [11], and both phases often coexist in ferroelectric $HfO_2$-based thin films [12,13]. Previous experimental and theoretical works have put their effort in identifying the mechanism(s) to stabilize the PO phase, which can be roughly classified into thermodynamic models (including various extrinsic factors such as doping, grain size effect, strain, and electric field) and kinetic models (accounting for phase transition barriers during the annealing process) [7,10-12,14-22]. However, a simple strategy to stabilize the PO phase remains elusive. Among the proposed mechanisms, the oxygen vacancy is brought into sharp focus for following reasons. First, oxygen vacancy is a dominant intrinsic defect in $HfO_2$ thin films which is unavoidable and often found at a relatively high concentration of $1.7 \times 10^{21}$ cm$^{-3}$ (corresponding to $HfO_{2-x}$ with x = ~2.65%) in experiments [23,24] . Second, there have been experimental observations that the M phase would transform to the PO phase when oxygen vacancies migrate from the interface into the bulk region driven by an electric field [23,25], and high oxygen vacancy concentration can suppress the M phase and stabilize the PO phase, so that significantly improve the ferroelectric property in $HfO_2$ films[26]. A very recent experimental work reported that ferroelectricity in $HfO_2$-based film is intertwined oxygen vacancy migration, and oxygen vacancy migration can induce structural phase transitions. Nevertheless, they claimed that the nature of coupling between oxygen vacancy and ferroelectric "remain agnostic" [27]. Theoretical studies have reported that the energy difference between the PO phase and the M phase will slightly decrease as the concentration of the charge neutral oxygen vacancy ($V_O$) increases; however, even at an extremely high concentration of ~12%, the energy of the PO phase is still substantially higher than that of the M phase [5,14,15,28]. It is evident that the oxygen vacancy alone cannot solve the stability puzzle. Moreover, first-principles calculations also indicate that the diffusion barrier of $V_O$ in $HfO_2$ is too high (~3 eV) to be mobile [29,30], which is also conflicting with the experimental observations [23,31].

This experiment-theory conundrum regarding oxygen vacancy in $HfO_2$ may derive from an additional degree of freedom: defect charge state. The charge states of a defect is a subject that has been widely studied in semiconductors; previous theoretical works focus on their formation energies, transition levels, etc. [32,33]. In particular, oxygen vacancies are known to possess different charge states such as the doubly positively charged state ($V_O^{2+}$) in various oxides including $HfO_2$[34,35]. To our best knowledge, no theoretical studies of the effect of $V_O^{2+}$ on the phase stability of oxides have yet been performed in $HfO_2$, similar idea have been performed in $ZrO_2$ by self-consistent tight-binding (SCTB) model[36]. Using density functional theory (DFT) calculations, we find that a $V_O^{2+}$ concentration of 2.35% is enough to make the PO phase thermodynamically more stable than the M phase in $HfO_2$. More important, we thoroughly study the effect of $V_O^{2+}$ on the relative phase stability in various wide-band-gap oxides. We propose that for wide-band-gap oxides, by comparing the absolute energy of defect levels induced by charge neutral oxygen vacancy in different competing phases, the relative thermodynamic stability between competing phases containing charged oxygen vacancies can be determined. Furthermore, we find that $V_O^{2+}$ exhibits lower diffusion barriers than that of $V_O$ in HfO$_2$, indicating the mobile oxygen vacancy observed in experiments is likely at +2 charge state. These results provide insight into the origin of ferroelectricity of HfO$_2$ and may help to explain the vacancy-induced nonpolar-polar phase transition in HfO$_2$.

*Methods*

The density functional theory (DFT) calculations are performed using a plane-wave basis set with a cutoff energy of 500 eV as implemented in the Vienna Ab initio Simulation Package (VASP)[37,38], and the electron exchange-correlation potential was described by the Perdew−Burke−Ernzerhof revised for solids scheme (PBEsol) of the generalized gradient approximation[39]. The PBE functional usually underestimate the band gap, our calculated band gap (~4 eV) is smaller than the experimental value of 5.9 eV[40]. To obtain more accurate information about the band gap, one may opt for hybrid exchange-correlation such as Heyd–Scuseria–Ernzerhof (HSE) functional[41]. The Brillouin zone of unit cell with 12 atoms is sampled with a 12 × 12 × 12 Monkhorst-Pack k-mesh. To perform the charged vacancy calculation based on DFT, a jellium background charge was used. The $V_O$ and $V_O^{2+}$ for both M and PO phase are modeled through a periodically repeated supercell of 2×3×2 (6×4×6 k-mesh) which contains 48 Hf and 96 O atoms. According to previous work, the electrostatic effects on energy of charged defects is inversely proportional to the inter-defect distance[42], thus the supercells studied in this work is sufficiently large to prevent coupling between periodic oxygen vacancies.

*Atomic and electronic structure*

The DFT calculations show that the lattice relaxation around $V_O$ is negligibly weak, while being remarkably strong around $V_O^{2+}$ in the M phase. For $V_O$ (removing a four-fold coordinated oxygen atom), the four neighboring Hf atoms are displaced outward by ~0.013 Å and neighboring O atoms are displaced inward by ~0.035 Å from their original lattice positions, corresponding to 0.6% and 1.6% of the equilibrium Hf-O bond length, respectively. In the case of $V_O^{2+}$, the neighboring Hf atoms have outward displacements of ~0.18 Å, and the surrounding O atoms are distorted inward by ~0.23 Å, as shown by blue and green arrows in Fig. 1a. The atomic displacements induced by $V_O^{2+}$ are about 8.6%–11.0% of the equilibrium Hf-O bond length, nearly ten times of the values around the charge neutral $V_O$. The displacements decay exponentially with increasing distance from the vacancy site (see Supplemental Material, Fig. S1 [43]). Similar effect of oxygen vacancy on lattice relaxations is also found in the PO phase that the local atomic distortions around $V_O^{2+}$ are much larger than those around $V_O$.

The markedly different lattice relaxations around $V_O$ and $V_O^{2+}$ result from the distinctions of their electronic structures. The removal of an oxygen atom leads to a doubly-occupied defect state within the band gap (see Fig. 1b) due to the wide band gap of HfO$_2$, while in narrow-band-gap oxides (such as SrTiO$_3$) the defect states will merge with the bottom of empty conduction band states, making the system metallic [44]. It is well known that semi-local density functionals such as Perdew-Burke-Ernzerhof (PBE) will underestimate the band gap. Considering that the PBE band gap of HfO$_2$ from our DFT calculations is already quite large (~4 eV), a further enhancement of the band gap (i.e., predicted with hybrid functionals such as HSE06) will only increase the tendency of electron localization around the vacancy site and the formation of doubly-occupied defect state within the band gap. Therefore, the presence of in-gap defect states is robust for wide-band-gap oxides [45] (We also perform calculations using HSE functional, see Supplemental Material Fig. S5 [43]). It is also worth noting that in Fig. 1b, we plot the density of states with respect to the absolute energy instead of setting the Fermi level to zero as usual; we also mark the difference between the defect levels in M phase and PO phase, which will be a focal point in our later discussions. The two excess electrons occupying the in-gap defect states are strongly localized at the oxygen vacancy site, forming an $F$-center type defect as shown in Fig. 1a. This unusual electronic structure is similar to the electride materials filled with localized electrons in crystal voids [46]. The two localized electrons act effectively as an O$^{2-}$ in the perfect crystal of HfO$_2$, therefore the neighboring atoms around $V_O$ can barely sense the removal of an oxygen atom and remain nearly undistorted. Conversely for $V_O^{2+}$, the neighboring atoms can sense the vacancy and the balance of lattice structure is broken, thus leading to strong lattice relaxations.

***Stabilizing the ferroelectric phase by charged oxygen vacancy***

To investigate the influence of charge state of oxygen vacancies on phase stability, we defined $\Delta E$ (eV per formula unit) as the total energy difference between PO phase and M phase as a function of the concentration of oxygen vacancies: $\Delta E(C_{Vo}^n) = E_{PO}(C_{Vo}^n) - E_M(C_{Vo}^n)$, where $C_{Vo}$ represents the concentration of oxygen vacancies, and superscript $n$ represents the charge state of vacancy, the $n = 0$ and $+2$ are used in calculations. In a 2×3×2 supercell with 96 O atoms, removing one oxygen atom corresponds to HfO$_{2-x}$ with $x = 2.08\%$. The PO and M phases are modeled with supercells of the same size containing the same number of atoms. This allows a direct comparison of the absolute energies between PO and M phases at the same charge state. A negative value of $\Delta E$ indicates PO phase is energetically more favorable than M phase.

The calculation results show that $V_O$ in HfO$_2$ cannot stabilize ferroelectric PO phase, because $\Delta E$ only slightly reduces but remains positive with the increasing concentration of $V_O$ as illustrated in Fig. 2a, which agrees well with previous theoretical studies [5,15,28]. For $V_O^{2+}$, as illustrated in Fig. 2a, $\Delta E$ decreases gradually from 36.7 meV/f.u. to −57.5 meV/f.u., as the concentration of $V_O^{2+}$ increases from 1.04 to 4.16 % (corresponding to 1 and 4 vacancies in 2×3×2 supercells). This indicates that $V_O^{2+}$ can stabilize the PO phase more effectively than $V_O$ and in good agreement with experimental observation that high oxygen vacancy concentration can suppress the M phase and stabilize the PO phase[26]. We observe a linear relationship between the concentration of $V_O^{2+}$ and $\Delta E$ (yellow line in Fig. 2a) at low concentrations ($C_{Vo}^{2+} <$ 2.0 %); the extrapolation gives a critical concentration of 2.35 % above which the PO phase is more stable thermodynamically than the M phase. This value agrees well with results from direct DFT calculations for high concentrations which are shown by the blue solid squares in Fig. 2a. The charge vacancy induced stabilization effect is presented even without atomic relaxations around the vacancy site (see blue dashed line in Fig. 2a). Therefore the ion relaxation is not the dominant factor responsible for the drastic reduction of $\Delta E$. The reduction of $\Delta E$ is mainly originated from the reduction of the electrostatic energy due to the removal of two localized electrons at the oxygen vacancy site, as discussed in detail below. Similar behavior is found for three-fold coordinated charged oxygen vacancies in other phases of $HfO_2$ (see Supplemental Material, Fig. S2 and S3 [43]).

In addition to the ferroelectric phase stability, the magnitude of polarization is rather crucial for device applications. We find $V_O^{2+}$ only slightly weakens the magnitude of the polarization of PO phase. In comparison, $V_O$ enhances the polarization. As shown in Fig. 2b, the polarization magnitude decreases with the increasing $V_O^{2+}$ concentration in PO phase. For instance, ~3 % of $V_O^{2+}$ will reduce the polarization by ~10% (from 52.2 to 46.7 μC/cm$^2$), but the resultant polarization magnitude remains within the acceptable range of electronic devices application. The suppressed polarization induced by charged oxygen vacancies might be one of the reasons why the experimental polarization value is usually lower than the ideal magnitude of 52.2 μC/cm$^2$ from first-principles calculation [1,4,7,23,25].

### *Structure phase stabilization influenced by charged oxygen vacancies in wide band gap oxides*

Considering that oxygen vacancy is common in oxides, we systematically examined the relationship between $V_O^{2+}$ and phase stability in a few representative oxides, and identify a general trend. We propose a thermodynamic cycle connecting the energy difference between two competing phases (*A* and *B* with *B* lower in energy) and the in-gap defect level energy difference (eq.1).

$$\Delta \varepsilon_D = \varepsilon_D^A - \varepsilon_D^B \tag{1}$$
$$E^A(V_o) = E^A(V_O^{2+}) + q\varepsilon_D^A \tag{2}$$
$$E^B(V_o) = E^B(V_O^{2+}) + q\varepsilon_D^B \tag{3}$$

It is noted that $\varepsilon_D$ is the energy of the defect level due to charge neutral oxygen vacancies, superscripts indicate that they correspond to phase *A* and *B*, as illustrated in the inset of Fig. 3. $E(V_o)$ and $E(V_O^{2+})$ in eq.2 and eq.3 indicate the total energy of specific phase with neutral or charged oxygen vacancy, respectively. We then define $\Delta E_P$, subtract eq.2 from eq.3 to get eq.4. The eq.4 is the relative phase stability in the presence of neutral $V_O$ and charged $V_O^{2+}$, respectively.

$$\Delta E_P = [E^A(V_o) - E^B(V_o)] - [E^A(V_O^{2+}) - E^B(V_O^{2+})] = q\Delta\varepsilon_D \tag{4}$$

Therefore, $\Delta E_P$ reflects the change in relative phase stability when the charge state of oxygen vacancy becomes +2. The key assumption in the thermodynamic cycle is that for wide-band-gap oxides, the energy of localized charge $q$ at a vacancy is simply $q\varepsilon_D$, and thus $\Delta E_P$ is equal to $q\Delta\varepsilon_D = 2\Delta\varepsilon_D$. A positive value of $\Delta E_P$ means the presence of $V_O^{2+}$ has the tendency to stabilize $A$ phase relative to the $B$ phase, potentially driving a phase transition from $B$ to $A$. It is evident that $V_O^{2+}$ will reduce more energy of $A$ phase if $\varepsilon_D^A > \varepsilon_D^B$. Take PO phase and M phase in HfO$_2$ as example, the energy of removing two excess electrons in PO phase is 1.08 eV higher than that in M phase. This value is approximately equal to twice the difference between defect levels which we marked in Fig. 1b (0.49 eV). As we discussed above, in wide gap oxides, the effective charge from $V_O$ is strongly localized at vacancy site and does not bond with atoms around the defect. Therefore, the energy of defect level is mainly contributed by the Hartree potential of the surrounding atoms, and the Hartree potential can be well described by DFT calculations. Due to the different arrangement of atoms around the defect in different phases, their Hartree potentials are also different, which is reflected in the difference of defect levels ($\Delta\varepsilon_D$). That's why for PO phase and M phase in HfO$_2$, $\Delta E_P$ is approximately equal to $q\Delta\varepsilon_D$.

The above conclusion about the relationship between $\Delta E_P$ and $q\Delta\varepsilon_D$ should be general for wide-band-gap oxides, and even for ionic insulator. Because Hartree potential and electrostatic energy is universal, and the important factor here is a large band gap that can form a localized defect level, thus the elemental composition and crystal structure of the material are less irrelevant. We first focus on hafnium dioxide and zirconium dioxide (ZrO$_2$), checking the $\Delta E_P$ of their various phases relative to the M phase. As shown in Figure 3, the coordinate of each dot is ($q\Delta\varepsilon_D$, $\Delta E_P$) and all dots are located around the dash line which represents $q\Delta\varepsilon_D = \Delta E_P$. We also calculate the oxygen vacancy in oxides of various valence states, of which band gap obtained by PBE is larger than 3.5 eV, as shown in Figure 3 (See Table S1 in Supplemental Material for detail data [43]), they also located around the dash line. Figure 3 indicates that $q\Delta\varepsilon_D = \Delta E_P$ is a universal phenomenon in wide band gap oxides. It means that for two specific phases of a certain wide band gap oxides, the $V_O^{2+}$ will significantly reduce the total energy of one phase relative to the other phase, and induce a phase transition. We notice that previous works focus on researching different defects in one certain system. The important parameter there is the chemical potential for the effective charge and atoms being removed or added [47-49]. The chemical potential of the effective charge is given by the chemical potential of the electrons, i.e. the Fermi energy which is usually referenced to the top of the valence band ($E_{VBM}$).[50,51] In our work, we are comparing two specific phases of one certain system; when comparing the energy difference between them, the term of chemical potential of atoms is cancelled. As for the energy of the electrons, it is not proper to set the $E_{VBM}$ of two phases as reference separately. Therefore, we uniquely select the defect levels that have been ignored in previous work as a reference.

### *Oxygen vacancy diffusion activation energy*

In ferroelectric HfO$_2$ films, it is often observed that oxygen vacancies migrate from the interface to the bulk region during an electric field cycling process [23,25]. The calculated oxygen vacancy diffusion activation energy ($E_a$) of M phase and PO phase are 2.89 and 2.49 eV for $V_O$, respectively, as shown in Fig. 4a, consistent with the value of 2.4−3.2 eV reported in previous work [29,52,53]. Such large diffusion barrier essentially prohibits the migration of $V_O$. Our calculations show that $V_O^{2+}$ has a much lower $E_a$ of 0.98 and 0.85 eV for M and PO phase respectively, as shown in Fig. 4a, which indicates that the mobile oxygen vacancy observed in experiments is likely at a charged state such as $V_O^{2+}$. The diffusion results of $E_a$ are shown to be agreement with those measured for positively charged oxygen vacancies with an activation energy of 0.52 eV in thin $HfO_2$ films [24], as well as DFT calculation activation energy barriers of 0.69 eV in M phase in previous work [54]. Similarly, the reason for the decrease in $E_a$ can be found in the electronic structure. As shown in Fig. 4b, for the initial configuration with a $V_O$, the two excess electrons, corresponding to the doubly-occupied defect state within the band gap, are strongly localized at the oxygen vacancy site. During the oxygen vacancy migration, the two localized electrons at the vacancy site hinder the migration of neighboring oxygen atom. At an intermediate state, the two localized electrons become delocalized and transfer to a defect state with higher energy, resulting in a higher $E_a$. For $V_O^{2+}$, the in-gap defect states are unoccupied and there is no effective charge at the oxygen vacancy site to oppose the incoming oxygen atom during the diffusion, and the $E_a$ is significantly lower. A similar trend has also been observed in $ZrO_2$ [55,56] and other oxides[57,58]

In summary, using first-principles DFT calculations, we systematically compared the effects of the $V_O^{2+}$ and $V_O$ on atomic/electronic structures, electric polarization, phase stability, and the activation energy of oxygen vacancy diffusion in $HfO_2$. As the concentration of $V_O^{2+}$ increases to greater than ~2 %, the ferroelectric PO phase becomes thermodynamically stable than the M phase, which indicates that $V_O^{2+}$ can help stabilize ferroelectric phase in $HfO_2$. We propose a general relationship between the energy in-gap defect levels and relative phase stability in the presence of $V_O^{2+}$ in wide-band-gap oxides. It is also found that compared to $V_O$, $V_O^{2+}$ has lower vacancy diffusion activation energy, likely responsible for the observed oxygen vacancy migrations in experiments. These results suggest that compared to the $V_O$, the theoretically predicted behavior of $V_O^{2+}$ is in better agreement with that observed in experiments. These results may be helpful for the understanding of the origin of ferroelectricity in $HfO_2$ and pave a new way for researching competing structural phases in wide-band-gap materials.

**ACKNOWLEDGMENTS**

This work was supported by the National Key R&D Program of China (Grant No. 2017YFA0303602), the Key Research Program of Frontier Sciences of CAS (Grant No. ZDBS-LY-SLH008), the National Nature Science Foundation of China (Grants No. 11774360, No. 11974365), and 3315 program of Ningbo. Calculations were performed at the Supercomputing Center of Ningbo Institute of Materials Technology and Engineering.

# References

[1]  T. S. Böscke, J. Müller, D. Bräuhaus, U. Schröder, and U. Böttger, Appl. Phys. Lett. **99**, 102903 (2011).

[2]  T. S. Böscke, S. Teichert, D. Bräuhaus, J. Müller, U. Schröder, U. Böttger, and T. Mikolajick, Appl. Phys. Lett. **99**, 112904 (2011).

[3]  J. Müller, U. Schröder, T. S. Böscke, I. Müller, U. Böttger, L. Wilde, J. Sundqvist, M. Lemberger, P. Kücher, T. Mikolajick, and L. Frey, J. Appl. Phys. **110**, 114113 (2011).

[4]  S. Mueller, J. Mueller, A. Singh, S. Riedel, J. Sundqvist, U. Schroeder, and T. Mikolajick, Adv. Funct. Mater. **22**, 2412 (2012).

[5]  H.-J. Lee, M. Lee, K. Lee, J. Jo, H. Yang, Y. Kim, S. C. Chae, U. Waghmare, and J. H. Lee, Science **369**, 1343 (2020).

[6]  S. S. Cheema, D. Kwon, N. Shanker, R. dos Reis, S.-L. Hsu, J. Xiao, H. Zhang, R. Wagner, A. Datar, M. R. McCarter, C. R. Serrao, A. K. Yadav, G. Karbasian, C.-H. Hsu, A. J. Tan, L.-C. Wang, V. Thakare, X. Zhang, A. Mehta, E. Karapetrova, R. V. Chopdekar, P. Shafer, E. Arenholz, C. Hu, R. Proksch, R. Ramesh, J. Ciston, and S. Salahuddin, Nature **580**, 478 (2020).

[7]  J. Müller, T. S. Böscke, U. Schröder, S. Mueller, D. Bräuhaus, U. Böttger, L. Frey, and T. Mikolajick, Nano Lett. **12**, 4318 (2012).

[8]  X. Sang, E. D. Grimley, T. Schenk, U. Schroeder, and J. M. LeBeau, Appl. Phys. Lett. **106**, 162905 (2015).

[9]  M. H. Park, Y. H. Lee, H. J. Kim, Y. J. Kim, T. Moon, K. D. Kim, J. Müller, A. Kersch, U. Schroeder, T. Mikolajick, and C. S. Hwang, Adv. Mater. **27**, 1811 (2015).

[10]  T. D. Huan, V. Sharma, G. A. Rossetti, and R. Ramprasad, Phys. Rev. B **90**, 064111 (2014).

[11]  R. Materlik, C. Künneth, and A. Kersch, J. Appl. Phys. **117**, 134109 (2015).

[12]  M. H. Park, T. Schenk, C. M. Fancher, E. D. Grimley, C. Zhou, C. Richter, J. M. LeBeau, J. L. Jones, T. Mikolajick, and U. Schroeder, J. Mater. Chem. C **5**, 4677 (2017).

[13]  E. D. Grimley, T. Schenk, T. Mikolajick, U. Schroeder, and J. M. LeBeau, Adv. Mater. Interfaces **5**, 1701258 (2018).

[14]  C. Künneth, R. Materlik, M. Falkowski, and A. Kersch, ACS Appl. Nano Mater. **1**, 254 (2018).

[15]  Y. Zhou, Y. K. Zhang, Q. Yang, J. Jiang, P. Fan, M. Liao, and Y. C. Zhou, Comput. Mater. Sci. **167**, 143 (2019).

[16]  R. Batra, T. D. Huan, J. L. Jones, G. Rossetti, and R. Ramprasad, J. Phys. Chem. C **121**, 4139 (2017).

[17]  P. Fan, Y. K. Zhang, Q. Yang, J. Jiang, L. M. Jiang, M. Liao, and Y. C. Zhou, J. Phys. Chem. C **123**, 21743 (2019).

[18]  M. H. Park, Y. H. Lee, H. J. Kim, T. Schenk, W. Lee, K. D. Kim, F. P. G. Fengler, T. Mikolajick, U. Schroeder, and C. S. Hwang, Nanoscale **9**, 9973 (2017).

[19]  Y. Qi, S. Singh, C. Lau, F.-T. Huang, X. Xu, F. J. Walker, C. H. Ahn, S.-W. Cheong, and K. M. Rabe, Phys. Rev. Lett. **125**, 257603 (2020).

[20]  S. E. Reyes-Lillo, K. F. Garrity, and K. M. Rabe, Phys. Rev. B **90**, 140103 (2014).

[21]  S. Liu and B. M. Hanrahan, Phys. Rev. Mater. **3**, 054404 (2019).

[22]  X. Xu, F.-T. Huang, Y. Qi, S. Singh, K. M. Rabe, D. Obeysekera, J. Yang, M.-W. Chu, and S.-W. Cheong, Nat. Mater. **20**, 826 (2021).

[23] M. Pešić, F. P. G. Fengler, L. Larcher, A. Padovani, T. Schenk, E. D. Grimley, X. Sang, J. M. LeBeau, S. Slesazeck, U. Schroeder, and T. Mikolajick, Adv. Funct. Mater. **26**, 4601 (2016).

[24] S. Zafar, H. Jagannathan, L. F. Edge, and D. Gupta, Appl. Phys. Lett. **98**, 152903 (2011).
[25] E. D. Grimley, T. Schenk, X. Sang, M. Pešić, U. Schroeder, T. Mikolajick, and J. M. LeBeau, Adv. Electron. Mater. **2**, 1600173 (2016).
[26] A. Pal, V. K. Narasimhan, S. Weeks, K. Littau, D. Pramanik, and T. Chiang, Appl. Phys. Lett. **110**, 022903 (2017).
[27] P. Nukala, M. Ahmadi, Y. Wei, S. De Graaf, E. Stylianidis, T. Chakrabortty, S. Matzen, H. W. Zandbergen, A. Björling, and D. Mannix, Science **372**, 630 (2021).
[28] M. Hoffmann, U. Schroeder, T. Schenk, T. Shimizu, H. Funakubo, O. Sakata, D. Pohl, M. Drescher, C. Adelmann, R. Materlik, A. Kersch, and T. Mikolajick, J. Appl. Phys. **118**, 072006 (2015).
[29] N. Capron, P. Broqvist, and A. Pasquarello, Appl. Phys. Lett. **91**, 192905 (2007).
[30] C. Tang, B. Tuttle, and R. Ramprasad, Phys. Rev. B **76**, 073306 (2007).
[31] S. Starschich, S. Menzel, and U. Böttger, Appl. Phys. Lett. **108**, 032903 (2016).
[32] C. Freysoldt, B. Grabowski, T. Hickel, J. Neugebauer, G. Kresse, A. Janotti, and C. G. Van de Walle, Rev. Mod. Phys. **86**, 253 (2014).
[33] S. T. Pantelides, Rev. Mod. Phys. **50**, 797 (1978).
[34] D. M. Ramo, A. Shluger, J. Gavartin, and G. Bersuker, Phys. Rev. Lett. **99**, 155504 (2007).
[35] D. M. Ramo, J. Gavartin, A. Shluger, and G. Bersuker, Phys. Rev. B **75**, 205336 (2007).
[36] S. Fabris, A. T. Paxton, and M. W. Finnis, Acta Mater. **50**, 5171 (2002).
[37] E. Cockayne, Phys. Rev. B **75**, 094103 (2007).
[38] G. Kresse and J. Furthmüller, Comput. Mater. Sci. **6**, 15 (1996).
[39] J. P. Perdew, A. Ruzsinszky, G. I. Csonka, O. A. Vydrov, G. E. Scuseria, L. A. Constantin, X. Zhou, and K. Burke, Phys. Rev. Lett. **100**, 136406 (2008).
[40] V. V. Afanas'ev, A. Stesmans, F. Chen, X. Shi, and S. A. Campbell, Appl. Phys. Lett. **81**, 1053 (2002).
[41] J. Heyd, G. E. Scuseria, and M. Ernzerhof, J. Chem. Phys. **118**, 8207 (2003).
[42] C. Freysoldt, J. Neugebauer, and C. G. Van de Walle, Phys. Rev. Lett. **102**, 016402 (2009).
[43] See Supplemental Material at link for Methods, Note, Table S1-S3 and Figure S1-S6, which includes Refs. [10,11,21,28,37-42,59-61].
[44] D. D. Cuong, B. Lee, K. M. Choi, H.-S. Ahn, S. Han, and J. Lee, Phys. Rev. Lett. **98**, 115503 (2007).
[45] A. Alkauskas, P. Broqvist, and A. Pasquarello, Phys. Rev. Lett. **101**, 046405 (2008).
[46] S. Matsuishi, Y. Toda, M. Miyakawa, K. Hayashi, T. Kamiya, M. Hirano, I. Tanaka, and H. Hosono, Science **301**, 626 (2003).
[47] S. Zhang and J. E. Northrup, Phys. Rev. Lett. **67**, 2339 (1991).
[48] C. G. Van de Walle, D. Laks, G. Neumark, and S. Pantelides, Phys. Rev. B **47**, 9425 (1993).
[49] S. Zhang, S.-H. Wei, and A. Zunger, Phys. Rev. B **63**, 075205 (2001).
[50] G. M. Dalpian and S.-H. Wei, Phys. Rev. Lett. **93**, 216401 (2004).
[51] G. M. Dalpian, Y. Yan, and S.-H. Wei, Appl. Phys. Lett. **89**, 011907 (2006).
[52] C. Tang, B. Tuttle, and R. Ramprasad, Phys. Rev. B **76**, 073306 (2007).
[53] Y. Dai, Z. Pan, F. Wang, and X. Li, AIP Advances **6**, 085209 (2016).
[54] N. Capron, P. Broqvist, and A. Pasquarello, Appl. Phys. Lett. **91**, 192905 (2007).
[55] A. S. Foster, V. Sulimov, F. L. Gejo, A. Shluger, and R. M. Nieminen, Phys. Rev. B **64**, 224108 (2001).
[56] J. Yang, M. Youssef, and B. Yildiz, Phys. Rev. B **97**, 024114 (2018).
[57] A. Kyrtsos, M. Matsubara, and E. Bellotti, Phys. Rev. B **95**, 245202 (2017).

[58] M. Y. Yang, K. Kamiya, B. Magyari-Köpe, M. Niwa, Y. Nishi, and K. Shiraishi, Appl. Phys. Lett. **103**, 093504 (2013).

[59] J. X. Zheng, G. Ceder, T. Maxisch, W. K. Chim, and W. K. Choi, Phys. Rev. B **75**, 104112 (2007).

[60] S. Clima, D. J. Wouters, C. Adelmann, T. Schenk, U. Schroeder, M. Jurczak, and G. Pourtois, Appl. Phys. Lett. **104**, 092906 (2014).

[61] G. Mills, H. Jónsson, and G. K. Schenter, Surf. Sci. **324**, 305 (1995).

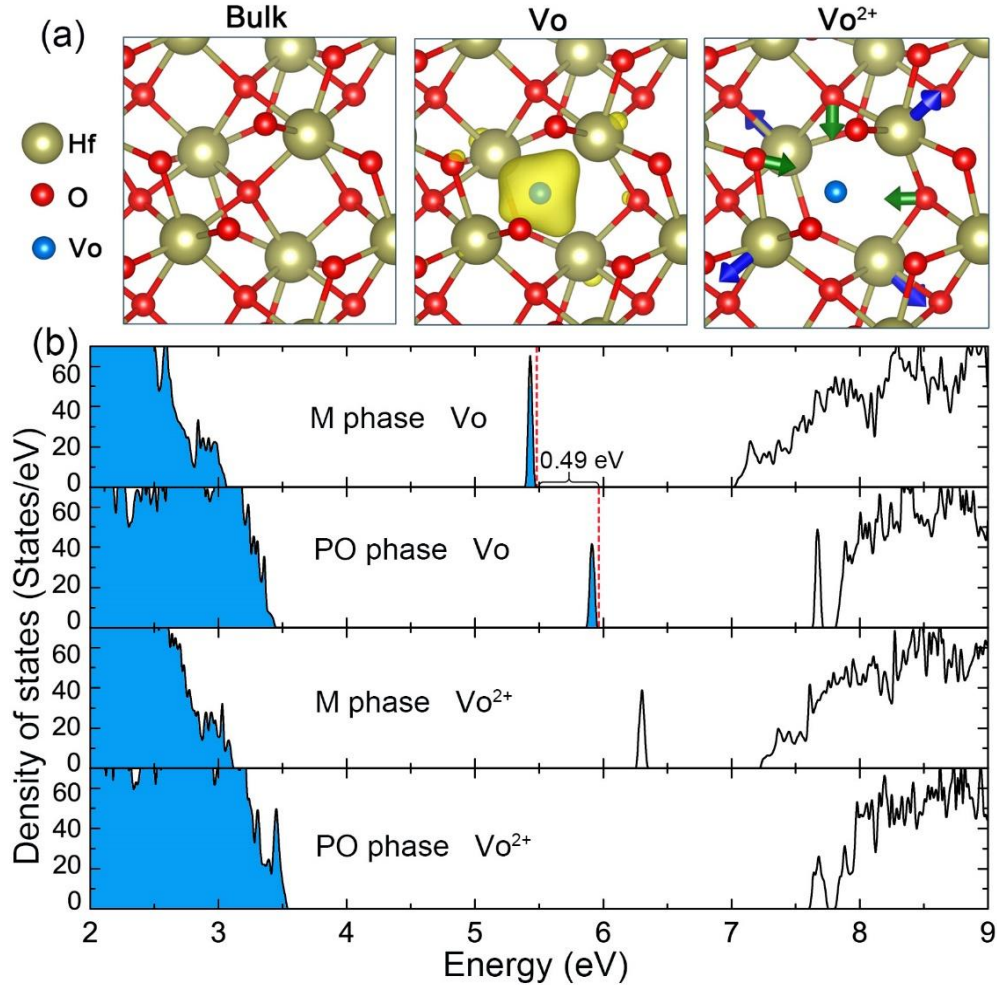

**Figure 1** (a) Defect-free atomic structure of M-phase HfO$_2$ and the local atomic relaxations around the charge neutral ($V_O$) and +2 charged ($V_O^{2+}$) oxygen vacancy; yellow surface represents the charge density isosurface of the doubly-occupied defect state due to charge neutral vacancy; the strong outward and inward displacements of neighboring Hf and O atoms around +2 charged oxygen vacancy are indicated by blue and green arrows. (b) Density of states of different phase of HfO$_2$ with different charge state vacancy, and blue shaded area represents occupied states. The values of doubly-occupied defect level is shown by red dash line. The energy difference of defect levels in M-phase and PO-phase is 0.49 eV.

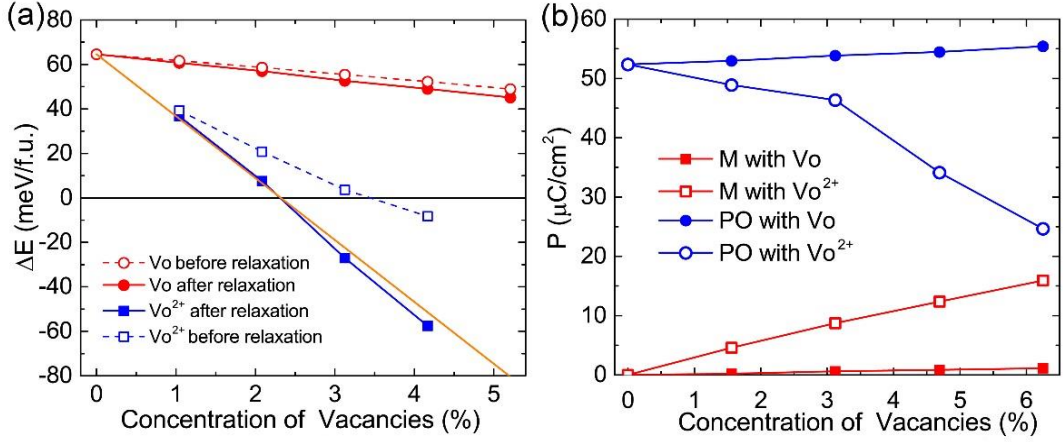

**Figure 2** (a) Energies of the PO phase of HfO$_2$ with different vacancy charge states and relaxation conditions relative to the M-phase ($\Delta E$) as a function of vacancy concentration obtained with a 2×3×2 supercells with 48 Hf and 96 O atoms. (b) Vacancy concentration dependence of local polarization for PO and M phases of HfO$_2$ with oxygen vacancy at different charge states.

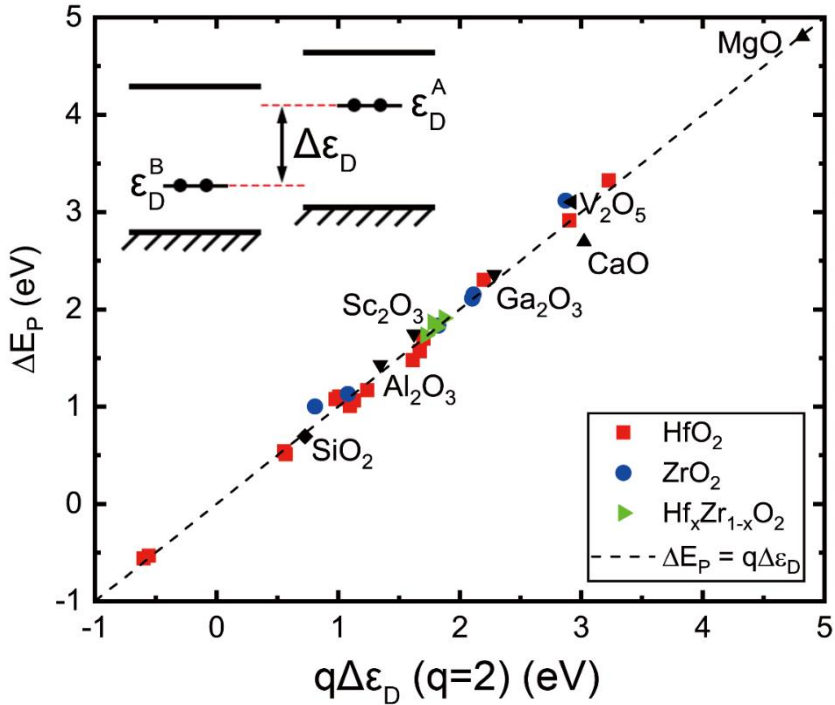

**Figure 3** The energy differences of defect levels multiplied by the effective charge, versus the energy differences between two phases in same formula cell of oxides in eq. 4.

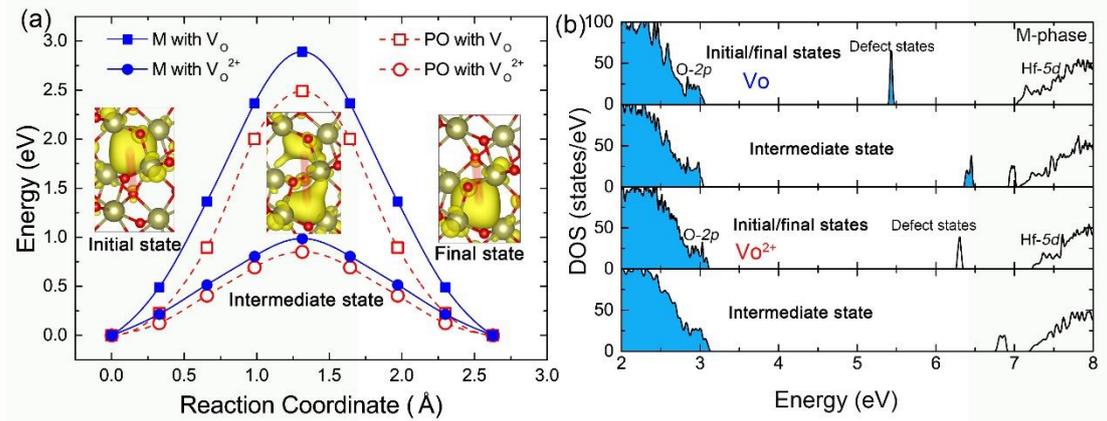

**Figure 4** (a) The energy variation of the $V_O$ and $V_O^{2+}$ during their migrations along the pathways shown in the inset and yellow surface represents the charge density isosurface of the defect state; (b) the corresponding density of states of initial/final and intermediate states with $V_O$ and $V_O^{2+}$ of M phase $HfO_2$.

# Supplemental Material

Ri He[1#], Hongyu Wu[1#], Shi Liu[2, 3, 4], Houfang Liu[5], X. Renshaw Wang[6, 7] and Zhicheng Zhong[1, 8*]

[1]Key Laboratory of Magnetic Materials Devices & Zhejiang Province Key Laboratory of Magnetic Materials and Application Technology, Ningbo Institute of Materials Technology and Engineering, Chinese Academy of Sciences, Ningbo 315201, China

[2]School of Science, Westlake University, Hangzhou, Zhejiang 310024, China

[3]Institute of Natural Sciences, Westlake Institute for Advanced Study, Hangzhou, Zhejiang 310024, China

[4]Key Laboratory for Quantum Materials of Zhejiang Province, Hangzhou Zhejiang 310024, China

[5]Institute of Microelectronics and Beijing National Research Center for Information Science and Technology (BNRist), Tsinghua University, Beijing 100084 China

[6]Division of Physics and Applied Physics, School of Physical and Mathematical Sciences, Nanyang Technological University, 21 Nanyang Link, 637371 Singapore

[7]School of Electrical and Electronic Engineering, Nanyang Technological University, 50 Nanyang Ave, 639798 Singapore

[8]China Center of Materials Science and Optoelectronics Engineering, University of Chinese Academy of Sciences, Beijing 100049, China

[#] These authors contribute equally to this work

[*] zhong@nimte.ac.cn

**Methods**

We consider three competing phases, including ground nonpolar monoclinic P2$_1$/c (M) phase, polar orthorhombic Pca2$_1$ (PO) phase and nonpolar orthorhombic Pbca (AO) phase in this work. The density functional theory (DFT) calculations are performed using a plane-wave basis set with a cutoff energy of 500 eV as implemented in the Vienna Ab initio Simulation Package (VASP)[1,2], and the electron exchange-correlation potential was described by the Perdew−Burke−Ernzerhof revised for solids scheme (PBEsol) of the generalized gradient approximation[3]. In order to further verify the accuracy of our calculation results, adequate plane-wave energy cutoffs were determined by examining the convergence of bulk. The total energy of a supercell of bulk was converged to 6 meV/f.u. at cutoff energy of 500 eV (relative to calculations at 600 eV), and see detail result in Table S3 in SM. Total energy and all forces on atoms converge to less than $1\times10^{-5}$ eV and 0.005 eV/Å, respectively. The PBE functional usually underestimate the band gap, our calculated band gap (~4 eV) is smaller than the experimental value of 5.9 eV[4]. To obtain more accurate information about the band gap, one may opt for hybrid exchange-correlation such as Heyd–Scuseria–Ernzerhof (HSE) functional[5]. We note that HSE functional only enlarge the band gap and formation of doubly-occupied defect state within the band gap, but the relative location of defect level in M and PO phase ($\Delta\varepsilon_D$) is 0.49 eV, being in good agreement with the PBE result (see Fig. 1b and Fig. S6a). We also have examined the results with local-density approximation (LDA) functional and found a similar $\Delta\varepsilon_D$ between M and PO phase (see Fig. S6b), on which the main conclusion of our work is based. Therefore, main conclusions in this work are not depend on the exchange-correlation functional in the calculations. The Brillouin zone of unit cell with 12 atoms is sampled with a 12 × 12 × 12 Monkhorst-Pack k-mesh. Table S2 lists the lattice parameters of the three phases of hafnia we used in this work, as predicted from our DFT calculations, which are in good agreement with experimental data and those from theoretical results reported before [6-9].

To perform the charged vacancy calculation base on DFT, we artificially set the number of electrons not equal with the number of the valence electrons of system, and assume a homogenous background charge to maintain the electrical neutrality of the periodic supercell, as known as the jellium background charge. If the number of electrons is less (greater) than the number derived from the valence and the number of atoms, the charge state is positive (negative), otherwise, the charge state is neutral. The neutral ($V_O$) and +2 charged oxygen vacancy ($V_O^{2+}$) for both M and PO phase are modeled through a periodically repeated supercell of 2×3×2 (6×4×6 k-mesh) which contains 48 Hf and 96 O atoms. According to previous work, the electrostatic effects on energy of charged defects is inversely proportional to the inter-defect distance[10], thus the supercells studied in this work is sufficiently large to prevent coupling between periodic oxygen vacancies. For comparison, we also adopt 2×2×2 supercells with 32 Hf and 64 O atoms in the calculation. All atomic positions and lattice constants are allowed to fully relax. The both threefold and fourfold coordinated oxygen vacancies are considered in the supercell models, because the stability of them strongly depends on the charge state of oxygen vacancy[11].

The polarization magnitude of unit cell was calculated by atomic displacements with respect to the referenced centrosymmetric P4$_2$/nmc tetragonal phase multiplied by the Born effective charges. The Born effective charges along the *c*-axis were calculated by DFT calculation: $Z_{Hf}^* = 4.86$, $Z_{O1}^* = -2.15$, $Z_{O2}^* = -2.61$, where O1 and O2 represent threefold and fourfold coordinated oxygens in HfO$_2$, the calculated spontaneous polarization of PO phase is 52.2 μC/cm$^2$, which agrees well with previous theoretical studies[12]. The activation energy ($E_a$) of oxygen vacancy migration between the nearest-neighbor sites was calculated by nudged elastic band (NEB) method[13].

We also calculate the oxygen vacancy in oxides of various valence states, these oxides are all common oxides, and their structure match experimentally determined crystal structure. The DFT method and calculation details of these oxides are consistent with that of HfO$_2$. The band gaps obtained by PBE of oxides listed in Table S1 are all larger than 3.5 eV. We repeat same calculation on the solid solutions of Hf$_x$Zr$_{1-x}$O$_2$ and strain condition of HfO$_2$ (modifying lattice parameters of a, b, c).

Since the defect states of $V_O$ and $V_O^{2+}$ are full-occupied and non-occupied, there is no spin-polarization in their electronic structure. We carried out the calculation of 1+ charged oxygen vacancy ($V_O^{1+}$), of which the defect state is occupied by one electron, the electronic structure shows a considerable spin-polarization induced by $V_O^{1+}$ (See Fig. S6).

| Phases of oxides | $\Delta\varepsilon_D = \varepsilon_D^A - \varepsilon_D^B$ (eV) | $\Delta E_P$ (eV) |
|---|---|---|
| **HfO$_2$** (B= P2$_1$/c; A= ) | | |
| Pca2$_1$ | 0.491 | 1.077 |
| Pca2$_1$ (3-coordination) | 0.359 | 0.848 |
| Pca2$_1$ (LDA) | 0.491 | 1.070 |
| Pbca | 0.506 | 1.105 |
| Pmna | 1.613 | 3.326 |
| R3 | 0.624 | 1.171 |
| R3m | 1.099 | 2.303 |
| P4$_2$/nmc | 0.860 | 1.700 |
| **HfO$_2$** (B= P2$_1$/c; A= P2$_1$/c with strain) | | |
| 3% | 0.8071 | 1.4802 |
| 2% | 0.5480 | 1.0055 |
| 1% | 0.2847 | 0.5125 |
| -1% | -0.2792 | -0.5329 |
| **HfO$_2$** (B=Pca2$_1$; A= Pca2$_1$ with strain) | | |
| 3% | 0.2777 | 0.5410 |
| 2% | 0.5645 | 1.0637 |
| 1% | 0.8348 | 1.5689 |
| -1% | -0.3011 | -0.5602 |
| **ZrO$_2$** (B= P2$_1$/c; A= ) | | |
| Pca2$_1$ | 0.541 | 1.131 |
| Pbca | 0.408 | 1.004 |
| R3 | 1.051 | 2.114 |
| R3m | 1.057 | 2.154 |
| P4$_2$/nmc | 0.912 | 1.836 |
| Fm$\bar{3}$m | 1.057 | 2.154 |
| **Hf$_x$Zr$_{1-x}$O$_2$** (B= P2$_1$/c; A= Pca2$_1$) | | |
| x=75% (Configuration 1) | 0.9053 | 1.8152 |
| x=75% (Configuration 2) | 0.8598 | 1.7327 |
| x=50% | 0.9352 | 1.9094 |
| x=25% | 0.8885 | 1.8666 |
| **Other binary oxides** | | |
| Al$_2$O$_3$ (A=Pbcn; B=R3c) | 0.6740 | 1.430 |
| Ga$_2$O$_3$ (A=C2/m; B=R3c) | 1.1435 | 2.358 |
| V$_2$O$_5$ (A= C2/c; B= P2$_1$/m) | 1.2715 | 3.214 |
| MgO (A=Fm$\bar{3}$m; B=P6$_3$mc) | 2.4095 | 4.805 |
| SiO$_2$ (A=P3$_2$21; B=P6$_2$22) | 0.3635 | 0.694 |
| CaO (A=Fm$\bar{3}$m; B=P6$_3$mc) | 1.5120 | 2.693 |
| Sc$_2$O$_3$ (A=Ia$\bar{3}$; B= P3m1) | 0.8118 | 1.747 |

**Table S1** The DFT calculated value of $\Delta\varepsilon_D$ and $\Delta E_P$ in HfO$_2$, ZrO$_2$, and other wide-band-gap binary oxides, the relationship between them satisfies the equation of $q \times \Delta\varepsilon_D \approx \Delta E_P$ ($q = 2$). The definition of $\Delta\varepsilon_D$ and $\Delta E_P$ are given in main text. The data are plotted out with colored dots in the Fig. 2.

| Phase | a (Å) | b (Å) | c (Å) | $\beta$(deg) | V(Å$^3$/u.f.) |
|---|---|---|---|---|---|
| P2$_1$/c (M) | 5.07 | 5.15 | 5.24 | 99.65 | 33.85 |
| Pca2$_1$ (PO) | 5.00 | 5.02 | 5.20 | 90 | 32.65 |
| Pbca (AO) | 9.94/2=4.97 | 5.02 | 5.19 | 90 | 32.45 |

**Table S2** Lattice parameters of optimized geometry of the monoclinic (M phase) and orthorhombic (PO and AO) phases of HfO$_2$

| P2$_1$/c Cutoff Energy | a (Å) | b (Å) | c (Å) | $\beta$ (deg) | V (Å$^3$/u.f.) | E (eV/f.u.) |
|---|---|---|---|---|---|---|
| 500 | 5.07 | 5.15 | 5.24 | 99.65 | -33.85 | -31.846 |
| 550 | 5.07 | 5.15 | 5.24 | 99.65 | -31.84 | -31.847 |
| 600 | 5.07 | 5.15 | 5.24 | 99.63 | -31.81 | -31.852 |
| 700 | 5.08 | 5.15 | 5.24 | 99.64 | -33.82 | -31.857 |

**Table S3** Lattice parameters and total energy of optimized geometry of the monoclinic (M phase) phase of HfO$_2$ with different cutoffs.

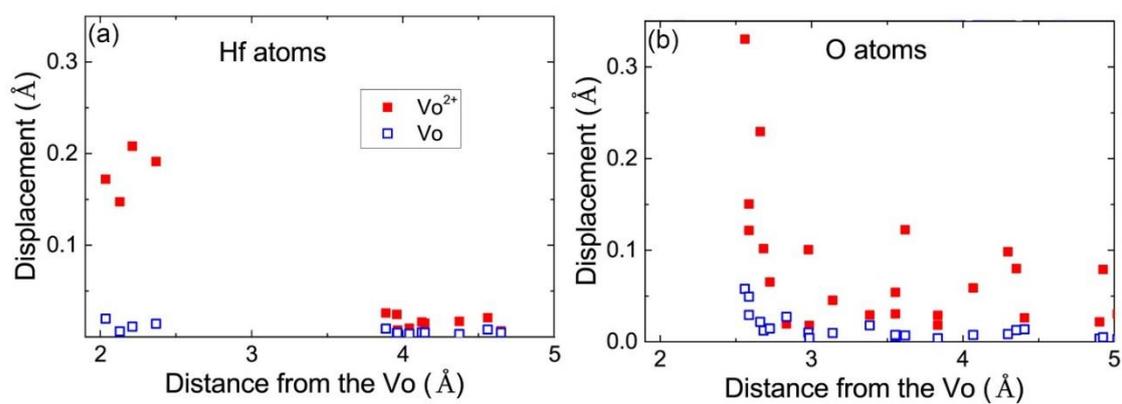

**Figure S1** The relaxation displacements with respect to original positon of (a) Hf and (b) O atoms in the supercells with $V_O$ and $V_O^{2+}$. The X-axis represents the distance between specific atom and vacancy site.

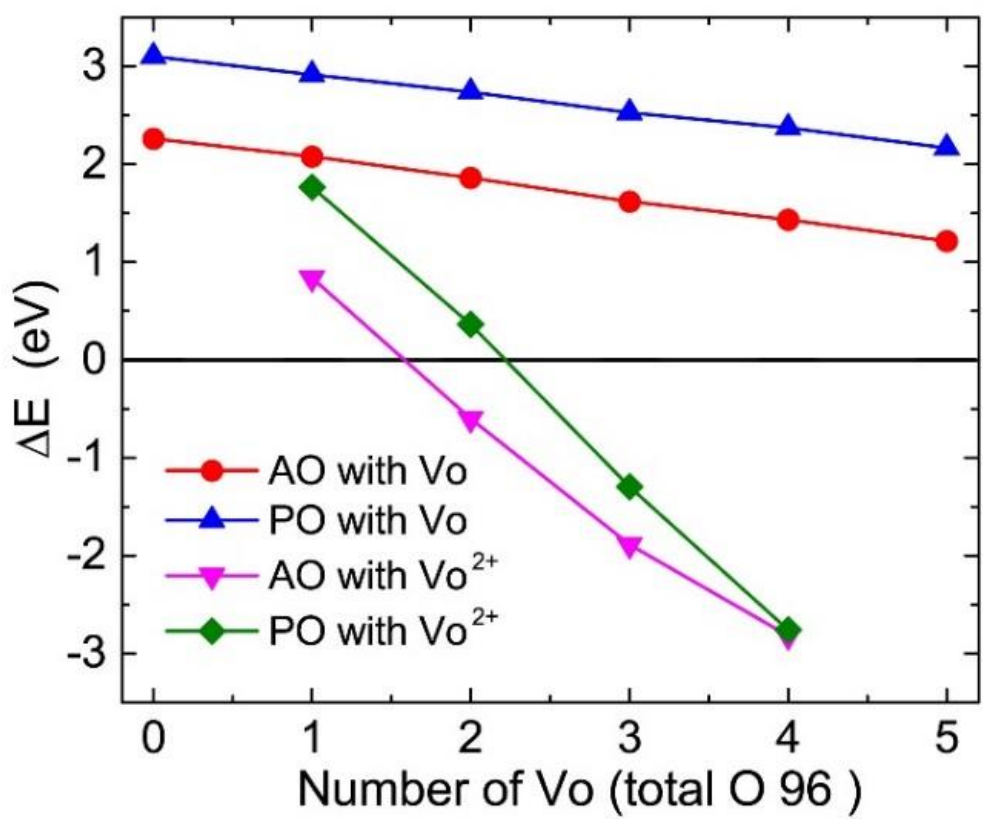

**Figure S2** The energies of AO phase and PO phase relative to that of the ground M-phase (ΔE) with $V_O$ and $V_O^{2+}$. The calculations are performed in a 2×3×2 supercells with 48 Hf and 96 O atoms.

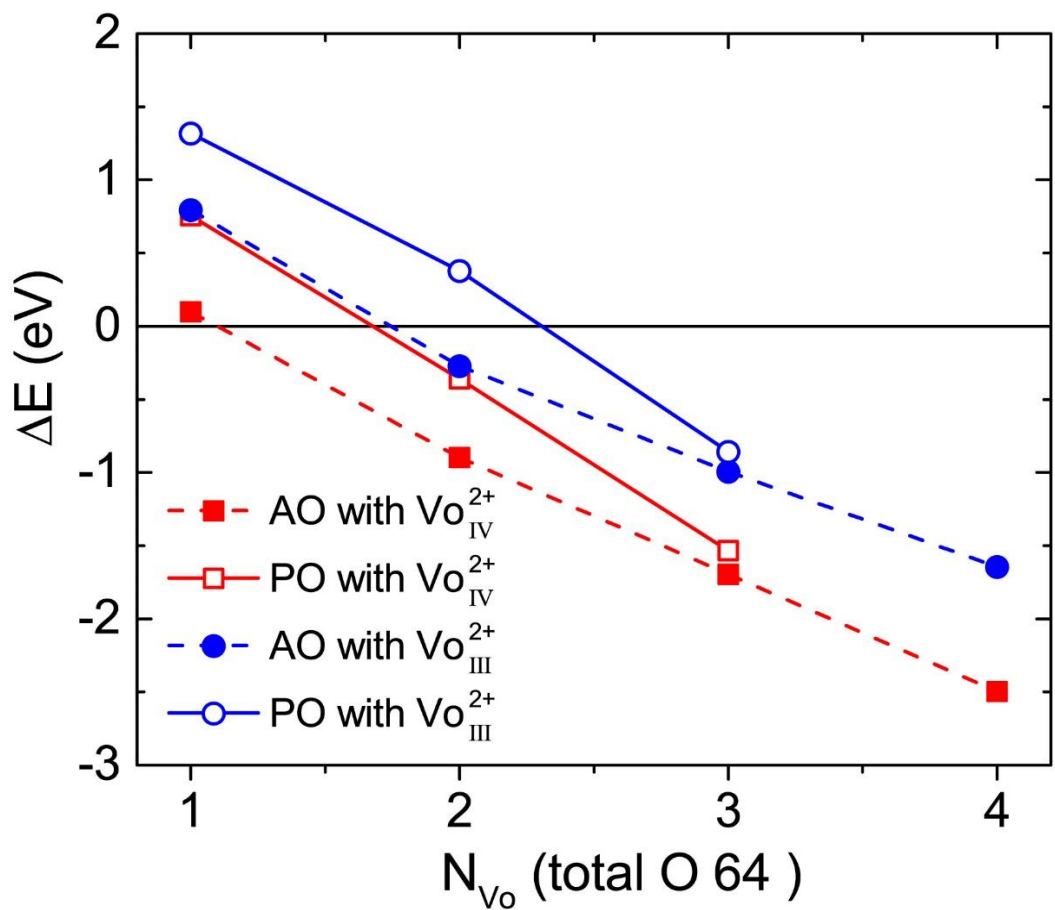

**Figure S3** The energies of AO phase and PO phase relative to that of the ground M-phase ($\Delta E$) with $V_O$ and $V_O^{2+}$, and with different coordination oxygen vacancy. The calculations are performed in a 2×2×2 supercells with 32 Hf and 48 O atoms.

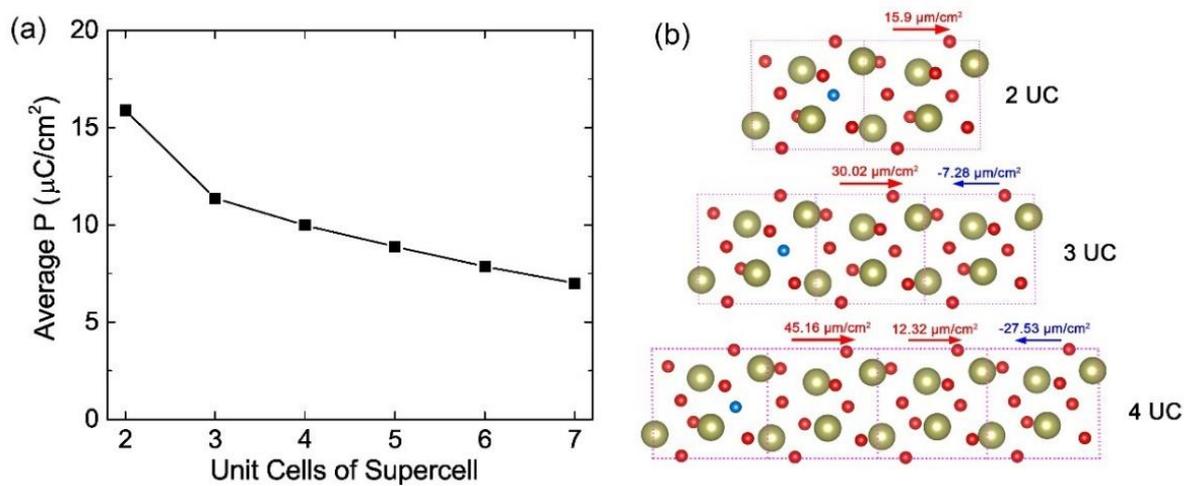

**Figure S4** (a) The total polarization for different size of supercells of M-phase $HfO_2$ and (b) the local polarization distribution induced by $V_O^{2+}$ in the supercells.

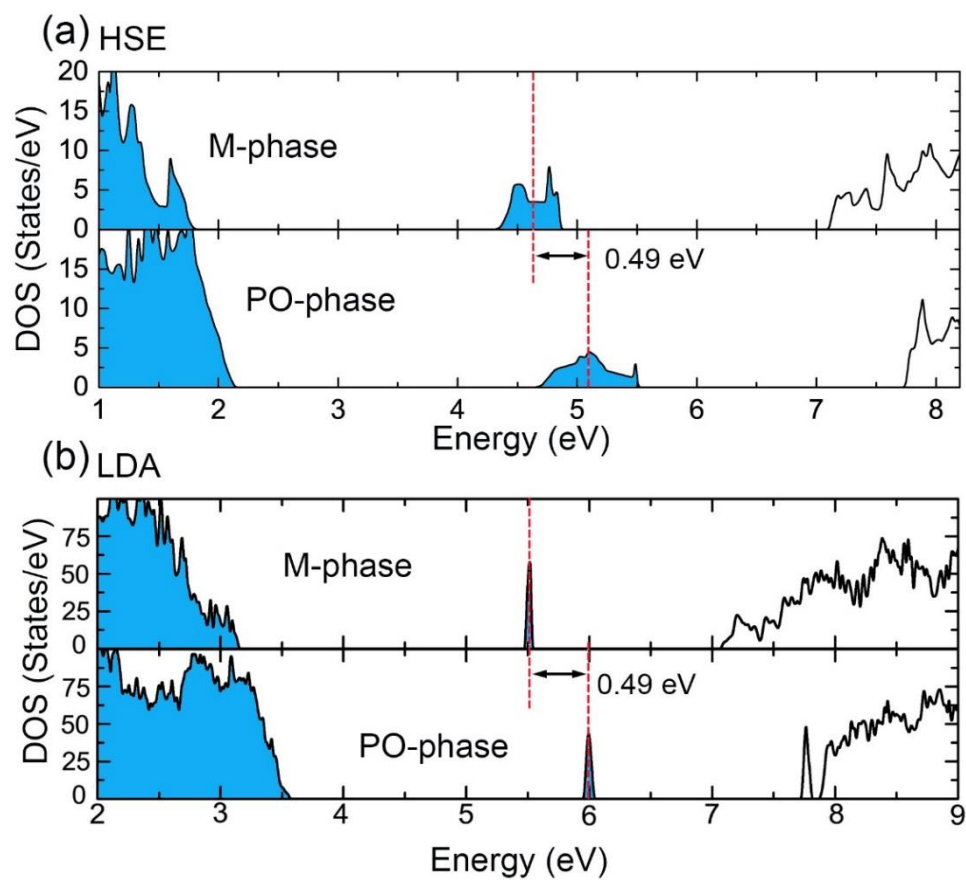

**Figure S5** Density of states (DOS) for M and PO phase of HfO$_2$ obtained using (a) HSE with 25% of exact exchange and (b) LDA. The red dish lines represent weighted average energy in defect states.

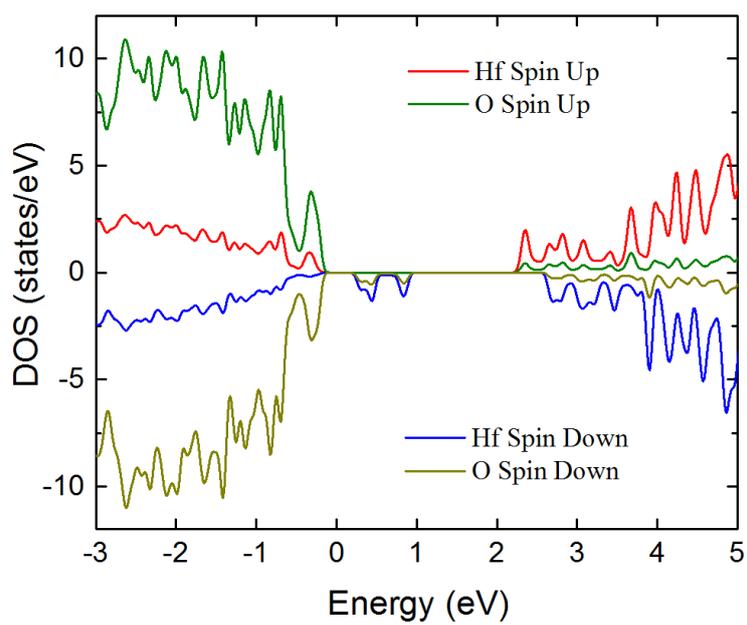

**Figure S6** Orbital-decomposed DOS of bulk HfO$_2$ with a +1 charged oxygen vacancy.

# References

[1] E. Cockayne, Phys. Rev. B **75**, 094103 (2007).

[2] G. Kresse and J. Furthmüller, Comput. Mater. Sci. **6**, 15 (1996).

[3] J. P. Perdew, A. Ruzsinszky, G. I. Csonka, O. A. Vydrov, G. E. Scuseria, L. A. Constantin, X. Zhou, and K. Burke, Phys. Rev. Lett. **100**, 136406 (2008).

[4] V. V. Afanas'ev, A. Stesmans, F. Chen, X. Shi, and S. A. Campbell, Appl. Phys. Lett. **81**, 1053 (2002).

[5] J. Heyd, G. E. Scuseria, and M. Ernzerhof, J. Chem. Phys. **118**, 8207 (2003).

[6] S. Liu and B. M. Hanrahan, Phys. Rev. Mater. **3**, 054404 (2019).

[7] T. D. Huan, V. Sharma, G. A. Rossetti, and R. Ramprasad, Phys. Rev. B **90**, 064111 (2014).

[8] R. Materlik, C. Künneth, and A. Kersch, J. Appl. Phys. **117**, 134109 (2015).

[9] M. Hoffmann, U. Schroeder, T. Schenk, T. Shimizu, H. Funakubo, O. Sakata, D. Pohl, M. Drescher, C. Adelmann, R. Materlik, A. Kersch, and T. Mikolajick, J. Appl. Phys. **118**, 072006 (2015).

[10] C. Freysoldt, J. Neugebauer, and C. G. Van de Walle, Phys. Rev. Lett. **102**, 016402 (2009).

[11] J. X. Zheng, G. Ceder, T. Maxisch, W. K. Chim, and W. K. Choi, Phys. Rev. B **75**, 104112 (2007).

[12] S. Clima, D. J. Wouters, C. Adelmann, T. Schenk, U. Schroeder, M. Jurczak, and G. Pourtois, Appl. Phys. Lett. **104**, 092906 (2014).

[13] G. Mills, H. Jónsson, and G. K. Schenter, Surf. Sci. **324**, 305 (1995).